\documentclass{elsarticle}
\usepackage{amsmath}
\usepackage{amssymb}
\usepackage{epsf}
\usepackage{graphicx}

\begin{document}

\title{Influence of interlayer asymmetry on magneto-spectroscopy of bilayer graphene}

\author[lancs]{M. Mucha-Kruczy\'{n}ski}

\author[lancs]{E. McCann\corref{cor1}}
\ead{e.mccann1@lancaster.ac.uk}

\author[lancs]{V.I. Fal'ko}

\cortext[cor1]{Corresponding author}
\address[lancs]{Department of Physics, Lancaster University,
Lancaster, LA1 4YB, UK}

\begin{abstract}

We present a self-consistent calculation of the interlayer
asymmetry in bilayer graphene caused by an applied electric field
in magnetic fields. We show how this asymmetry influences the
Landau level spectrum in bilayer graphene and the observable
inter-Landau level transitions when they are studied as a function
of high magnetic field at fixed filling factor as measured
experimentally in Ref.~\cite{kim_cr_bilayer}. We also analyze
the magneto-optical spectra of bilayer flakes in the photon-energy
range corresponding to transitions between degenerate and split
bands of bilayers.

\end{abstract}


\maketitle

\section{Introduction}
\label{sec:intro}

Bilayer graphene \cite{mccann_bilayer, novoselov_bilayer,
ohta_gap, oostinga_gap, mccann_bilayer_gap, min_gap, castro_gap}
is one of several graphite allotropes \cite{geim_review} that
display unique physical properties determined by the hexagonal
symmetry of their two-dimensional crystalline structure. The
electronic band structure of bilayer graphene close to the Fermi
energy consists of two degenerate bands touching each other at the
charge neutrality point (the position of the Fermi energy in a
neutral system) and two bands split by the interlayer coupling
\cite{mccann_bilayer, ohta_gap, rotenberg_arpes_prl}. In
bilayer-based field-effect transistors a band gap in the
electronic structure can be opened using a transverse electric
field that breaks the symmetry between the layers \cite{ohta_gap,
mccann_bilayer_gap, min_gap, castro_gap}, which may play an
important role in forming the transport characteristics of such
devices \cite{oostinga_gap}.

In this paper we analyze self-consistently the interlayer
asymmetry parameter for bilayer graphene field-effect transistors,
using the tight-binding approximation. The focus of this study is
the manifestation of interlayer asymmetry in the
magnetospectroscopy of bilayers, and we perform a self-consistent
analysis for bilayer flakes subjected to quantizing magnetic
fields, taking into account the possibility that there is finite
asymmetry in a neutral structure.
When the material is kept at constant filling factor upon the
variation of magnetic field (the measurement scheme employed in
recent experiment \cite{kim_cr_bilayer}), the necessity to vary
the charge density on the layers causes significant asymmetry
and the Landau level (LL) spectrum in a strong magnetic field is
altered considerably. Our calculation generalizes the
self-consistent analysis \cite{mccann_bilayer_gap} developed for
bilayers at zero magnetic field, and the analysis presented here
improves the rigour of Ref.~\cite{castro_ll} where the variation
of interlayer asymmetry on density and its influence on Landau
level transition energies was estimated by neglecting screening
effects. Also, we calculate magneto-optical spectra of bilayers
in the infra-red spectral range that has recently become
accessible in optical experiments using cyclotron irradiation
sources \cite{kim_optics_bilayer, fogler_optics_bilayer,
kuzmenko_optics_bilayer}, and where transitions between degenerate
and split bands of the bilayer may occur.

\section{The Landau level spectrum in charged bilayers: self-consistent analysis}
\label{sec:ll_spectrum}

\begin{figure}[t]\label{fig1}
\begin{center}
\includegraphics[width=0.7\hsize]{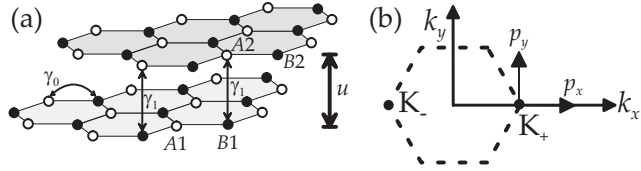}
\caption{(a) Schematic view of the bilayer graphene lattice
containing four atoms in the unit cell: A1 (white circles) and B1
(black) in the bottom layer and A2 (white) and B2 (black) in the
top one. Also shown are the hopping parameters $\gamma_{0}$ and
$\gamma_{1}$ used in the tight-binding model and interlayer
asymmetry $u$. (b) First Brillouin zone of bilayer graphene with
two inequivalent $K$ points: $K_{+}$ and $K_{-}$. Note the
difference between two momentum frames of reference:
$\mathbf{k}=\mathbf{K}_{\xi}+\mathbf{p}$ where
$\mathbf{k}=(k_{x},k_{y})$ and $\mathbf{p}=(p_{x},p_{y})$.}
\end{center}
\end{figure}

A schematic view of bilayer graphene (marked with the hopping
integrals considered throughout this paper) and the Brillouin zone
of bilayer graphene are shown in Fig. 1(a) and 1(b), respectively.
Bilayer graphene consists of two coupled hexagonal lattices with
inequivalent sites $A1$, $B1$ and $A2$, $B2$ in the first and
second graphene sheets, respectively, which are arranged according
to Bernal ($A2$-$B1$) stacking \cite{mccann_bilayer}. The
hexagonal Brillouin zone has two inequivalent degeneracy points
$K_{\xi}=(\xi\frac{4\pi}{3a},0)$ (where $\xi\in\{+,-\}$ and $a$ is
the lattice constant). Here, we take into account only the
nearest-neighbour in-plane and interlayer ($A2$-$B1$) coupling and the
lowest-order terms in the electron band energy expansion in
electron momentum (determined as the deviation of the electron
wave number from the corner $K_{+}$ or $K_{-}$ of the hexagonal
Brillouin zone that is identified below as the centre of the valley),
which corresponds to the Hamiltonian:
\begin{equation}\label{Hamilt}
{\hat{H}} \!\!= \! \!\xi\left(
\begin{array}{cccc}
-\frac{u}{2} & 0 & 0 & v\pi^{\dagger} \\
0 & \frac{u}{2} & v\pi & 0 \\
0 & v\pi^{\dagger} & \frac{u}{2} & \xi\gamma _{1} \\
v\pi & 0 & \xi\gamma_{1} & -\frac{u}{2}
\end{array}
\right).
\end{equation}
It is written in a basis of sublattice Bloch states
$\Psi_{+} = ( \psi_{A1}, \psi_{B2}, \psi_{A2}, \psi_{B1} )^{T}$ in
valley $K_{+}$ and
$\Psi_{-} = (\psi_{B2}, \psi_{A1}, \psi_{B1}, \psi_{A2})^{T}$ in
valley $K_{-}$, and $v$ is related to the nearest-neighbour hopping
parameter $\gamma_{0}$ ($v=\frac{a\sqrt{3}}{2\hbar}\gamma_{0}$)
and $\pi=p_{x}+i p_{y}$. The interlayer-asymmetry parameter $u$,
which will be at the heart of this study, describes the
on-site energy difference between different layers in the bilayer.

In the presence of a magnetic field $B$ perpendicular to the
flake, the bilayer spectrum splits into Landau levels (LL). The LL
spectrum can be obtained from the Hamiltonian in Eq.
(\ref{Hamilt}) using the Landau gauge $\mathbf{A}=(0,-Bx)$ and the
fact that operators $\pi=p_{x}+i p_{y}-eBx$ and
$\pi^{\dag}=p_{x}-i p_{y}+eBx$ coincide with lowering (raising)
operators in the basis of Landau functions $\psi_{m}(x,y)=e^{i
q_{y}y/\hbar}\phi_{m}(x+q_{y}\lambda_{B}^{2})$ (where
$\phi_{m}(x)$ are the wave functions of a quantum harmonic
oscillator),
\begin{equation}\label{pi_op}
\pi\psi_{m}=-i\frac{\hbar}{\lambda_{B}}\sqrt{2m}\psi_{m-1},
~\pi\psi_{0}=0,~and~\pi^{\dagger}\psi_{m}=i\frac{\hbar}{\lambda_{B}}\sqrt{2(m+1)}\psi_{m+1},
\end{equation}
where $\lambda_{B}=\sqrt{\hbar/eB}$ stands for magnetic length.

For a symmetric bilayer, $u=0$, energies $\epsilon_{m}$ of the
Landau levels are described by \cite{mccann_bilayer, abergel_ll,
cuniberti_ll, koshino_ll, nakamura_ll}:
\begin{eqnarray}\label{symmetric_c}
&\epsilon_{0}&=0;\nonumber \\ &\epsilon_{m\beta}^{c(s)}&=\frac{\beta}{\sqrt{2}}
\left(
\gamma_{1}^{2} + \Gamma(2m-1) \pm \sqrt{\gamma_{1}^{4}+2\Gamma\gamma_{1}^{2}(2m-1)+\Gamma^{2}}
\right)^{\frac{1}{2}},m\geq 1;
\end{eqnarray}
where $\Gamma=2\frac{\hbar^{2}v^{2}}{\lambda_{B}^{2}}$ and $\beta$
denotes conduction band ($+$) and valence band ($-$) LLs. Indices
$c$ and $s$ stand for low and high-energy (split) bands and
correspond to minus and plus signs in front of the square root,
respectively.

Nonzero asymmetry $u$, caused by a possible difference in electric potential
energy between the layers, modifies the LL spectrum
\cite{mccann_bilayer, guinea_ll, pereira_ll, misumi_ll,
mucha-kruczynski_ll}. To model this effect, we employ a
self-consistent theory of the charging of bilayer graphene. In
order to reproduce the conditions of recent experiments
\cite{kim_cr_bilayer} where the transition energy between
low-energy LLs was measured as a function of a varying magnetic
field but for a fixed filling factor, $\nu$, we extend the
self-consistent analysis of Ref. \cite{mccann_bilayer_gap} from
the zero-magnetic field regime into the regime of quantizing
magnetic fields, taking into account the possibility that there is
finite asymmetry in a neutral structure [see Eq.~(\ref{gap_eq})
below]. The analysis presented here improves the rigour of
Ref.~\cite{castro_ll} where the variation of interlayer asymmetry
on density and its influence on Landau level transition energies
was estimated by neglecting screening effects.

In particular, we consider a gated bilayer with interlayer
separation $c_{0}$. In external magnetic field $B$, a total excess
density, $n=\nu\frac{eB}{h}$, must be induced using the gate in
order to keep the filling factor $\nu$ fixed while changing $B$.
The density $n$ is shared between the two layers: $n=n_{1}+n_{2}$
where $n_{1}$ ($n_{2}$) is the excess density on the layer closest
to (furthest from) the gate. The difference in electric potential
between the layers is related to an incomplete screening of the
gate electric field by the charge $en_{1}$ on the first layer
alone and can be related to the unscreened density $n_{2}$,
\begin{equation}\label{gap_eq}
u(\nu,B)=w+\frac{e^{2}c_{0}n_{2}(\nu,B)}{\epsilon_{0}\epsilon_{r}}.
\end{equation}
Here $\epsilon_{r}$ is the effective dielectric constant
determined by the $\mathrm{SiO_{2}}$ substrate, and $w$ takes into
account finite asymmetry of a neutral structure (internal electric
field due to, for example, initial non-intentional doping of the
flake by deposits/adsorbates). In our numerical calculations we
use $\epsilon_{r} = 2$.

On the one hand, $u$ influences the LL spectrum via the
Hamiltonian in Eq.~(\ref{Hamilt}). On the other hand, its value depends
on the charge density $n_{2}$ which can only be obtained with a
full knowledge of the LL spectrum and the wave functions
corresponding to each LL. Therefore, a calculation of $u$ requires
a self-consistent numerical analysis. This calculation consists of
the following steps: for each given $B$, $5<B<20$T, and $\nu$ we
choose a starting $u$, and diagonalize Hamiltonian Eq.~(\ref{Hamilt})
to find the LL spectrum and the eigenstates with $m\lesssim M_{max}$
where $M_{max} \sim 300$.
Then, we sum over all filled Landau levels and determine the
excess electron densities on each layer. Note that, as a nonzero
value of $u$ splits the valley degeneracy of the LLs
\cite{mucha-kruczynski_ll}, care has to be taken when comparing
densities in specific LLs in biased and neutral structures, not to
confuse levels in different valleys. Finally, using Eq.
(\ref{gap_eq}) we find the asymmetry parameter and, then, iterate
the numerical procedure to obtain the self-consistent value of $u$
\cite{footnote_1}. Note that, for a sufficiently large cutoff
$M_{max} \sim 300$, the results were independent of $M_{max}$.

\begin{figure}[t]\label{fig2}
\begin{center}
\includegraphics[width=\hsize]{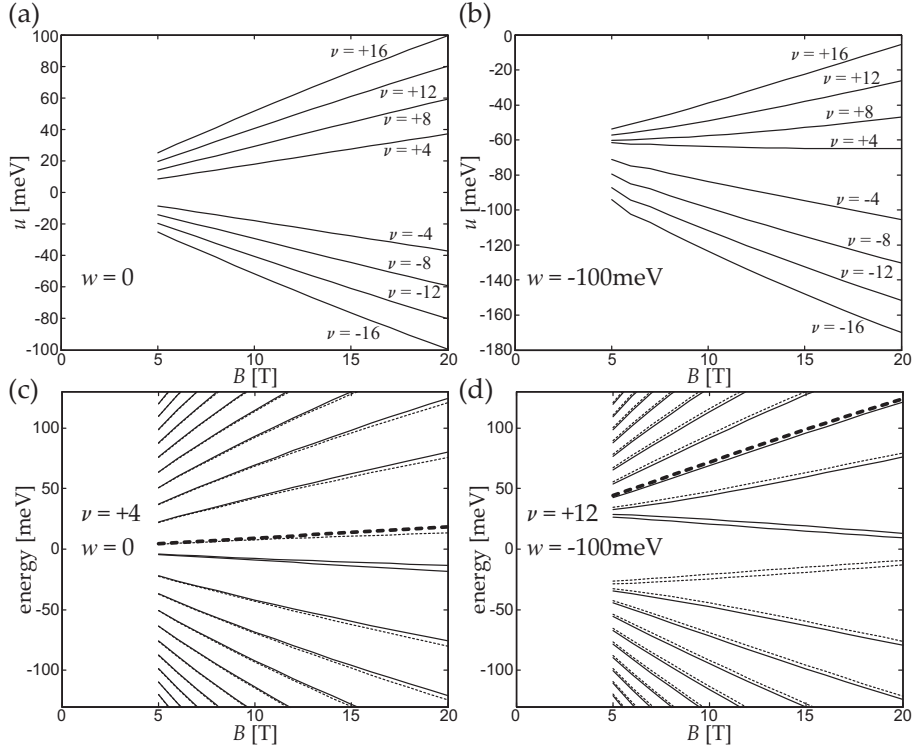}
\caption{Top row: results of a self-consistent calculation of the
interlayer asymmetry $u$ for (a) $w=0$ and (b) $w=-100$meV. Bottom
row: the LL spectrum as a function of applied magnetic field $B$
for constant filling factor and excess density-induced interlayer
asymmetry $u$: (c) $\nu=+4$; $w=0$, and (d) $\nu=+12$;
$w=-100$meV. Solid and dashed lines denote levels belonging to
$K_{+}$ and $K_{-}$, respectively. The line corresponding to the
last filled Landau level is shown in bold. In these calculations
we used $v=10^6$m/s and $\gamma_{1}=0.4$eV.}
\end{center}
\end{figure}

The self-consistently calculated values of $u$ obtained for several values
of the filling factor $\nu$ are shown in Fig. 2(a) and 2(b) for
the case of $w=0$ and a nominal $w=-100$meV, respectively. In the
case when $w=0$, the induced interlayer asymmetry is antisymmetric
with respect to the change of the filling factor from positive to
negative. This is because changing the filling factor from $+\nu$
to $-\nu$ corresponds to reversing the applied electric field and
inducing excess densities $-n$,$-n_{1}$ and $-n_{2}$ and thus
reversing the sign of $u$. Also, with decreasing $B$ all curves
tend towards $u=0$ and $u\approx-60$meV in Fig. 2(a) and 2(b),
respectively. These values are the results of the self-consistent
calculation with corresponding values of $w$ in the absence of a
magnetic field \cite{mccann_bilayer_gap}. Examples of the
low-energy LL spectrum for $\nu=+4$, $w=0$ and for $\nu=+12$,
$w=-100$meV are shown in Fig. 2(c) and 2(d). To list the LLs in
Fig. 2, we use three symbols: $sm\xi$, where $s$ attributes the LL
to the conduction (+) or valence (-) band, $m$ is the LL index and
$\xi\in(+,-)$ identifies the valley ($K_{+}$ or $K_{-}$) that the
level belongs to, respectively. The Landau levels $m=0,1$ have no
$s$ index, as those levels are degenerate when $u=0$
\cite{mccann_bilayer, mucha-kruczynski_ll}. The sign of the valley
splitting of the level $sm$ depends on the sign of $u$: for $u>0$,
level $sm+$ has higher energy than level $sm-$ whereas the
opposite is true for $u<0$. Levels $m=0,1$ behave differently: in
this case the energy $\epsilon_{m+(-)}<(>)0$ if $u>0$ and
$\epsilon_{m+(-)}>(<)0$ if $u<0$. The size of the valley splitting
of the low-energy LLs increases with $u$ and $B$ and for
$|u|\approx 0.1$eV, $B\approx 20$T [filling factors $\nu=+12,+16$
in Fig. 1(a) and $\nu=-8,-12,-16$ in Fig. 1(b)] is of the order of
10meV.

\section{Low-energy inter-Landau level transitions and bilayer signature in the FIR absorption}
\label{sec:ir_cr}

\begin{figure}[t]\label{fig3}
\begin{center}
\includegraphics[width=\hsize]{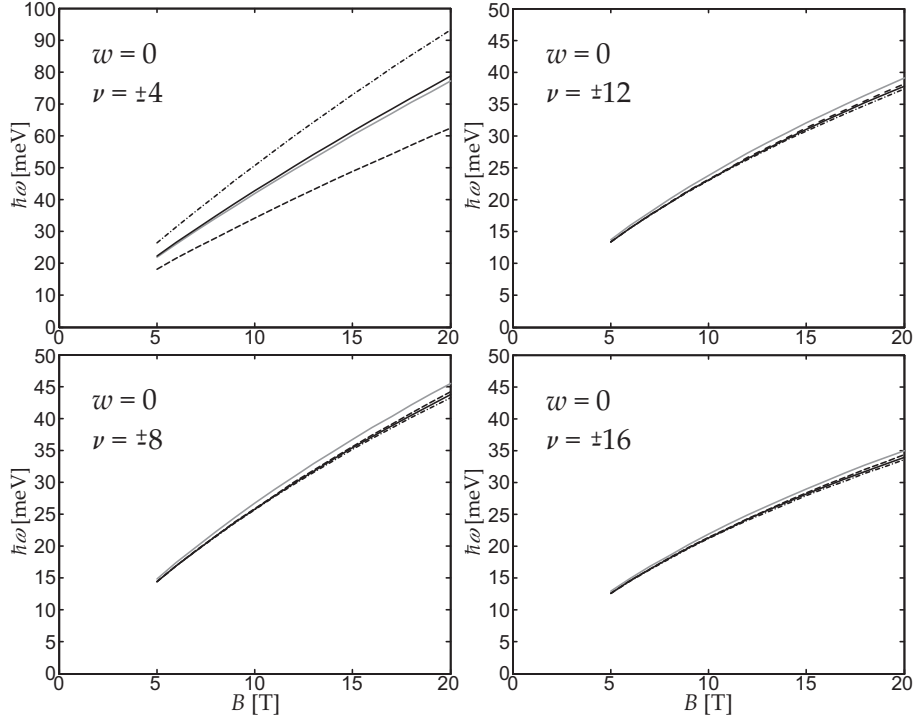}
\caption{Energy of low-energy inter-LL excitations as a function of
magnetic field for $w=0$. The broken lines are the contributions
of individual valleys to the transition energy:
black dot-dashed and dashed lines denote the
transition energy for positive (negative) $\nu$ at $K_{+}$
($K_{-}$) and $K_{-}$ ($K_{+}$), respectively. The solid black lines
show the contribution of both valleys to the transition energy,
calculated according to Eq.~(\ref{int}) (in this case the transition
energy is the same for both positive and negative $\nu$), whereas solid grey
lines depict the transition energy in a neutral ($u=0$) structure.
Note that for $\nu=8,12,16$ all black lines are very close to each
other and difficult to resolve.}
\end{center}
\end{figure}

Using spectra similar to those shown in Fig. 2(c) and 2(d), we
find the energy of the low-energy inter-Landau level transitions
for several filling factors which mimics the experimental
conditions of Ref.~\cite{kim_cr_bilayer} (the tight-binding
approach to this problem has also been adopted in Ref.~\cite{castro_ll}
where the dependence of interlayer asymmetry on
density $n$ and its influence on Landau level transition energies
was estimated by neglecting screening effects). In that
experiment, infrared light of energy $\hbar\omega$ was incident on
the graphene bilayer in a strong external magnetic field and with
a constant filling factor in order to excite charge carriers
between a prescribed pair of LLs and to probe the energy dispersion.
According to the selection rules determined in Ref.
\cite{abergel_ll} (and later extended for $u\neq 0$
\cite{mucha-kruczynski_ll}), only transitions which change the LL
index $m$ by one are allowed. Also, as photons provide a very
small momentum transfer, we only consider transitions between
levels at the same $K$ point. Thus, the corresponding
low-energy transitions for filling factors $\nu=+4,+8,+12,+16$ are
$1\xi\rightarrow +2\xi$, $+2\xi\rightarrow +3\xi$,
$+3\xi\rightarrow +4\xi$, $+4\xi\rightarrow +5\xi$ and for
$\nu=-4,-8,-12,-16$ they are $-2\xi\rightarrow 1\xi$,
$-3\xi\rightarrow -2\xi$, $-4\xi\rightarrow -3\xi$,
$-5\xi\rightarrow -4\xi$, respectively. However, as transitions
between the same levels at different $K$ points differ too little
in energy to have been resolved separately in the abovementioned
experiment (in fact, they can be only be clearly distinguished in
Fig. 3 for the case $\nu=4$), we obtain a single transition energy
$\epsilon_{trans}^{\nu}$ for a given filling factor $\nu$ by
comparing the relative intensities of the corresponding transition
at each $K$ point:

\begin{figure}[t]\label{fig4}
\begin{center}
\includegraphics[width=\hsize]{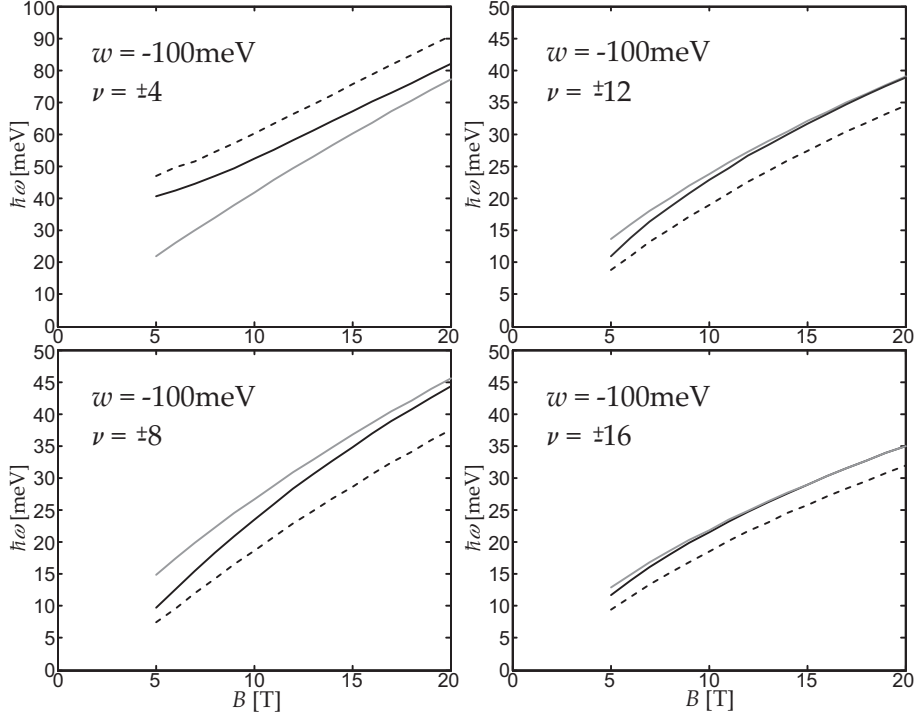}
\caption{Energy of low-energy inter-LL excitations as a function of
magnetic field for $w=-100$meV. Black solid and dashed lines
denote the transition energy for positive and negative filling
factor, respectively. Grey solid lines depict the transition energy in
a neutral ($u=0$) structure.}
\end{center}
\end{figure}

\begin{equation}\label{int}
\epsilon_{trans}^{\nu}=\frac{\epsilon_{trans}^{\nu}(K_{+})I^{\nu}(K_{+})
+
\epsilon_{trans}^{\nu}(K_{-})I^{\nu}(K_{-})}{I^{\nu}(K_{+})+I^{\nu}(K_{-})}
\end{equation}
where $\epsilon_{trans}^{\nu}(K_{\xi})$ and $I^{\nu}(K_{\xi})$ are
the transition energy at the $K_{\xi}$ point and its intensity,
respectively. The results obtained for sets of parameters as in
Fig. 2(a) and (b) are shown in Fig. 3 ($w=0$) and Fig. 4
($w=-100$meV). We shall first discuss the case $w=0$ presented in
Fig. 3. In this case, for a specified value of $B$, the asymmetry
$u$ changes sign with a change of the sign of the filling factor
[Fig. 2(a)], the Landau level spectrum for $\nu$ and $-\nu$ are the
same but the $K$ points have to be exchanged. Therefore,
$\epsilon_{trans}^{\nu}(K_{+})=\epsilon_{trans}^{-\nu}(K_{-})$,
clearly seen in all four graphs. Moreover, both transitions have
the same intensity and contribute equally to
$\epsilon_{trans}^{\nu}$ (black solid line). Comparison with the
transition energy for a neutral bilayer (grey line) shows
that non-zero $u$ decreases the energy of the transition. The
greater $|u|$ and $B$, the bigger the difference between
excitation energy in symmetric and biased bilayers. However, this
difference decreases with an increase of filling factor.

Introduction of parameter $w$ breaks symmetry between the
conduction and valence band LLs as presented in Fig. 4 for the
case of $w=-100$meV. The valence band excitation has greater
energy than the conduction band excitation at filling factor
$\nu=\pm 4$. However, this situation is reversed for higher
filling factors $\nu=\pm 8,\pm 12,\pm 16$ (this reversal was not observed in the experiment \cite{kim_cr_bilayer}). For this specific case, $w=-100$meV, the asymmetry introduced between excitations for filling
factors $\nu$ and $-\nu$ is of the size of $3-10$meV. These two
effects, the reduction of the transition energy with the increase
of $u$ and the breaking of the symmetry between transitions for
positive and negative filling factor caused by $w$, partly account
for the disagreement between experimental findings and
Eq.~(\ref{symmetric_c}) obtained from a tight-binding model for
neutral bilayers as used in Ref.~\cite{kim_cr_bilayer} to fit the
data. Other
investigations \cite{kusminskiy_ee, abergel_ee} show that additional
corrections may arise from electron-electron interactions.

\section{IR magneto-optics in bilayer graphene}
\label{sec:magnetooptics}

\begin{figure}[t]\label{fig5}
\begin{center}
\includegraphics[width=0.65\hsize]{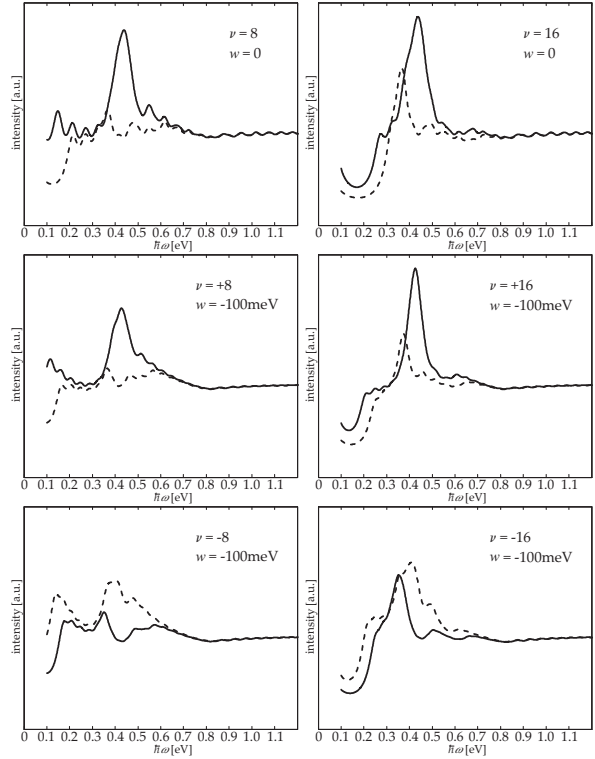}
\caption{Magneto-optical absorption spectra for bilayer graphene
in strong external magnetic field $B=14$T and for filling factors
$\nu=8$ and $\nu=16$ (left and right columns, respectively) and the
case of $w=0$ (top row) and $w=-100$meV (middle and bottom row).
For the symmetric case of $w=0$, solid and dashed lines show
absorption of right-handed (left-handed) and left-handed
(right-handed) circularly polarised light for the positive
(negative) filling factor, respectively. For the case of
$w=-100$meV, solid and dashed lines represent absorption of right
and left-handed circularly polarised light, respectively.}
\end{center}
\end{figure}

In this section, we analyse the optical transition spectra corresponding to
transitions between LLs in split bands of the bilayers. Here,
we use the self-consistently calculated interlayer-asymmetry
parameter $u$, and the LL energies and wave functions, and we
compute the infrared optical absorption spectra \cite{abergel_ll,
nicol_optical} of right ($\oplus$) and left-handed ($\ominus$)
circularly polarized light
$\mathbf{E}_{\omega}=E_{\omega}\mathbf{l}_{\oplus/\ominus}$, with
$\mathbf{l}_{\oplus}=(\mathbf{l}_{x}-i\mathbf{l}_{y})/\sqrt{2}$
and
$\mathbf{l}_{\ominus}=(\mathbf{l}_{x}+i\mathbf{l}_{y})/\sqrt{2}$
for bilayer graphene in a strong external magnetic field. The
broadening of the Landau levels is modeled using a Lorentzian
shape with the same full width at half maximum $\gamma=60$meV for
all Landau levels. Numerical results for magnetic field
$B=14$T and filling factors $\nu=8$ and $\nu=16$ are shown in
Fig. 5. For the case of $w=0$, the symmetry of the system demands
that the intensity of absorption of light with a given
polarisation for filling factor $\nu$ and at the $K_{\xi}$ point
is the same as that of the light with the inverted polarisation at
the $K_{-\xi}$ point for filling factor $-\nu$. This, indeed, is
the case for graphs in the left column of Fig. 5, where black
solid and dashed lines show absorption of right-handed
(left-handed) and left-handed (right-handed) circularly polarised
light for the positive (negative) filling factor, respectively.
Such a symmetry is broken for the case of $w=-100$meV, for which
the spectra for positive and negative filling factors are shown
in the separate panels, where solid and dashed
lines refer to right-handed and left-handed polarisation,
respectively. In particular, the peak visible for some of the
spectra at the radiation energy around $0.4$eV corresponds to
electron excitation between the low-energy $m=0$ LL and one of the
two high-energy $m=1$ LLs. Its position can be used to determine
the value of the coupling constant $\gamma_{1}$, and a small
shift in energy of this peak is due to strong magnetic field and
asymmetry $u$. Presented curves show a similar shape as those
predicted theoretically \cite{abergel_ll, nicol_optical} and
observed experimentally \cite{kim_optics_bilayer,
fogler_optics_bilayer} for IR optical absorption in the biased
bilayer.

\section{Summary}
\label{sec:summary}

In this work, we have considered gated bilayer graphene in
external magnetic field. We have shown that keeping the filling
factor constant results in a breaking of the symmetry between the
graphene layers due to the induced carrier density. We have
calculated the interlayer asymmetry $u$ self-consistently and
demonstrated its influence on the Landau level spectrum. Using
these results, we discussed both low-energy and high-energy
inter-LL excitations and compared them to recent experiments
concerning optical absorption. In particular, we achieved some
improvement over the standard tight-binding model for neutral
bilayer in the explanation of the cyclotron resonance experiment
probing low-energy the Landau level spectrum.

\section{Acknowledgements}
\label{sec:ack}

The authors would like to thank A. H. Castro Neto, E. A. Henriksen,
and S. Viola Kusminskiy for discussions. MM-K and EM would like to thank T.
Ando and the Tokyo Institute of Technology for hospitality. This
project has been funded by EPSRC Portfolio Partnership
EP/C511743/1, ESF CRP ``SpiCo,'' EPSRC First Grant EP/E063519/1,
and the Daiwa Anglo-Japanese Foundation.


\begin{thebibliography}{99}
\bibitem{footnote_1} Neither Landau level broadening
nor partial filling factors were considered in this work.
\end{thebibliography}
\end{document}